\documentclass{pnastwo}
\usepackage[dvips]{graphicx}
\usepackage{amssymb,amsfonts,amsmath}

\contributor{Edited by Joshua Jortner, Tel Aviv University, Tel Aviv, Israel, and approved June 24, 2009 (received for review April 20, 2009)}
\copyrightyear{2009}
\issuedate{August 18, 2009}
\volume{106}
\issuenumber{33}

\begin{document}






\title{Survival of an evasive prey}




\author{G. Oshanin\affil{1}{Laboratory J.-V. Poncelet (UMI  CNRS 2615),
Independent University of Moscow, Bolshoy Vlasyevskiy Pereulok 11, 119002 Moscow, Russia}\affil{2}{Laboratoire de Physique Th\'eorique de la Mati\`ere Condens\'ee (UMR CNRS 7600), Universit\'e Pierre et Marie Curie, 4
place Jussieu, 75252 Paris Cedex 5 France}\thanks{To whom
correspondence should be addressed. E-mail:
oshanin@lptmc.jussieu.fr}, O. Vasilyev\affil{3}{Max-Planck-Institut f\"ur Metallforschung,
Heisenbergstr. 3, D-70569 Stuttgart, Germany and
Institut f\"ur Theoretische und Angewandte Physik,
University of Stuttgart, D-70569 Stuttgart, Germany},
P. L. Krapivsky\affil{4}{Department of Physics, Boston University, Boston,
MA 02215 USA}, \and J. Klafter\affil{5}{School of Chemistry, Tel Aviv University, 69978 Tel Aviv, Israel}\affil{6}{Freiburg Institute for Advanced Studies (FRIAS), University of Freiburg, 79104 Freiburg, Germany}
}



\maketitle

\begin{article}

\begin{abstract}
We study the survival of a prey that is hunted by $N$ predators.
The predators perform independent random walks on a square lattice with $V$ sites
and start a direct chase
whenever the prey appears within their sighting range.
The prey is caught when a predator jumps to the site occupied by the prey.
We analyze the efficacy of a lazy, minimal-effort evasion strategy
according to which the prey tries to avoid encounters with the predators by making a hop only when any of the predators appears within its sighting range; otherwise the prey stays still. We show that if the sighting range of such a lazy prey is equal to one
lattice spacing, at least three predators are needed in order to catch the prey on a square lattice. In this situation, we establish a simple asymptotic
relation $\ln P_{\rm ev}(t)\sim (N/V)^2 \ln P_{\rm imm}(t)$ between the survival probabilities
of an evasive and an immobile prey.
Hence, when the density $\rho = N/V$ of the predators is low, $\rho \ll 1$, the lazy evasion strategy
leads to the spectacular increase of the survival probability. We also
argue that a short-sighting prey (its sighting range is smaller than the sighting range of the predators) undergoes an effective superdiffusive motion, as a result of its encounters with the predators, whereas a far-sighting prey performs a diffusive-type motion.
\end{abstract}




\section{Introduction and statement of results}

\dropcap{P}ursuit-and-evasion problems
have a long and fascinating history \cite{nahin}. The classical setup involves two agents --- say, a merchant vessel pursued by a pirate ship that it desperately tries to evade. The goal for both is
to choose a deterministic motion strategy, given their velocities and sighting ranges, that optimizes their respective chances of
successful pursuit or evasion.
Similar games between adversary species
occur in different environmental or biological systems;
co-evolution of bacteria and phage
or prey-predators contests being just two examples
(see, e.g., \cite{webb,miller,pascal0} for more details).
The chief difference here is that the objects move
less deterministically, their strategies are less ``intelligent'' and
the number of interacting objects can be large. Such
pursuer-evader contests are also assisted by some finite-range vision or smell.

\begin{figure}[ht]
  \centerline{\includegraphics*[width=0.26\textwidth]{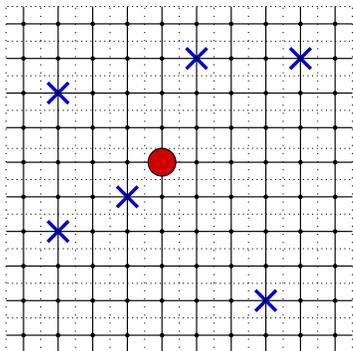}}
\caption{A prey (a circle) and predators (crosses) on a square lattice with $V$ sites.}
  \label{cartoon1}
\end{figure}

In this paper we investigate a class of pursuit-and-evasion problems involving a single evading prey that is being hunted by $N$ predators  (Fig.~\ref{cartoon1}) - a variation of the classic princess-and-monster game \cite{isaacs}.
The predators perform \textit{independent} nearest-neighbor random walks (RWs)
on a finite square lattice (with $V$ sites),
and the prey is caught upon the first encounter with a predator.
In such a situation, how should the prey move in order to maximize its chances of not being caught up to time $t$?

When the prey is blind, i.e. it has no information on predators' actual positions (its sighting range is zero), the best strategy is to stay still \cite{pascal2,pascal1}.
The survival probability $P_{\rm imm}(t)$ of this immobile prey is given by
\begin{equation}
\label{survival2}
P_{\rm imm}(t) \sim e^{- \alpha \, \rho \, t}\,,
\end{equation}
where $\rho = N/V$ and $\alpha$ depends on the diffusion coefficient of the predators, the structure of the lattice (particularly, the dimensionality),
and on the number of sites $V$.

The exponential decay \eqref{survival2} is an ultimate asymptotic for a finite lattice. There is also an intermediate asymptotic (which is the true asymptotic
for infinitely large systems) \cite{target}
\begin{equation}
\label{survival}
P_{\rm imm}(t) = e^{- \rho \, S(t)}\,,
\end{equation}
where $S(t)$ is the mean number of distinct sites visited
by a predator up to time $t$.

The behavior of $S(t)$ crucially depends on the spatial dimension $d$.
For nearest-neighbor RWs, $S(t)$ grows according to \cite{dvor}
\begin{equation}
\label{sites}
 S(t) \sim \left(\frac{8 t}{\pi}\right)^{1/2} \,\, {\rm and} \,\,\, S(t) \sim \frac{\pi t}{\ln(t)},
\end{equation}
on one-dimensional ($1d$) and two-dimensional ($2d$) square lattice, respectively. In three (and higher) dimensions,  $S(t)$ grows linearly with time, $S(t) \sim t/G$, where $G$ is the mean number of visits to the origin on an infinite lattice within an infinite time. (Hereinafter the symbol ``$\sim$'' signifies
that we deal with the leading in time asymptotic behavior.)



The question we address is how the simplest evasion strategy
affects the survival probability of a prey, having a finite sighting range $r$,
hunted by $N$ predators with sighting ranges $R$. The predators perform independent random walks and, as soon as a given predator appears within distance $R$ from the prey, it changes the mode of motion and
begins a direct chase, minimizing at each step a distance to the prey.
The prey tries to avoid encounters with the predators by investing a minimal effort:
Because the blind prey's best recourse is to stay still,
we assume that the lazy prey does the same as long as all the predators are outside
of its sighting range.
Whenever predators appear within its sighting range, the prey instantaneously hops to the nearest-neighboring site, chosen at random but so that (i) the distance from the visible predators will increase, and (ii) no other predator will get inside the prey's sighting range. We emphasize that whenever the prey hops,
the choice of the landing site is random modulo the validity of the above requirements.

We focus on the simplest situation when the sighting ranges of the lazy prey
and of the predators are both equal to just one lattice spacing. We find that in this case,
instead of obeying \eqref{survival2}--\eqref{survival},  the prey survival probability obeys:
\begin{equation}
\label{survival3}
P_{\rm ev}(t) \sim
\begin{cases}
e^{- B\alpha \, \rho^3 \, t} &{\rm finite \, lattice}\\
e^{- B\rho^3 \, S(t)}  &{\rm infinite \, lattice}
\end{cases}
\end{equation}
where $B$ is a numerical factor. The key feature is the replacement of the density $\rho$ by $\rho^3$. A rough explanation is that on the square lattice at least three predators must surround the prey in order to catch it. Hence, a very modest investment in effort pays back with a spectacular (several orders of magnitude)
increase of the survival probability.

We also observe that, due to encounters with the predators, the prey performs long-ranged excursions on the lattice until the moment when it is captured.
When $(r,R)=(1,1)$, the motion of the prey is effectively a diffusive motion with the mean-squared displacement growing linearly with time and the diffusion coefficients dependent on the mean density of predators.
 Surprisingly enough, there is a qualitative change in the behavior when $R>r$; we show numerically that in the case  $(r,R)=(1,2)$, the mean-squared displacement grows as $t^{1.65}$, i.e. random motion of the prey is
superdiffusive.

\section{The model and numerical results}

The minimal model is defined as follows:
\begin{enumerate}
\item There are $N$ predators on a square lattice with $V$ sites and periodic boundary conditions. Predators are placed at random and the prey is initially at the origin of the lattice.
\item Each predator performs a nearest-neighbor RW, the updates are made simultaneously, and the predators do not interact, e.g., there is no exclusion implying that a few predators can be at the same lattice site.
\item If a predator is on the nearest-neighboring site to the prey, it hops to that site and thereby the prey is caught.
\item Just before the predators hop, the prey checks the nearest-neighboring sites. If all four of them are empty, the lazy prey remains at the site. If some are occupied, the prey hops to the randomly chosen unoccupied nearest-neighboring site provided all four sites neighboring  the target one are unoccupied. If the latter happens, the predators chasing the prey still hop to the site that has been occupied by the prey.
\end{enumerate}

Thus, for the minimal model, the sighting ranges of both prey and predators are equal to one lattice spacing, $r=R=1$. The generalization is obvious: The predator starts a direct chase when the distance to the prey is $\leq R$; the prey begins to move whenever there are predators within the range $\leq r$.  By definition, the metric is Manhattan, so e.g. the site $(x,y)$ is on distance $|x|+|y|$ from the origin. We focus on the minimal model. (A few simulation results for the model with $r=1, R=2$ are presented below on  Fig.~\ref{cartoon5}.)

\begin{figure}[ht]
  \centerline{\includegraphics*[width=0.4\textwidth]{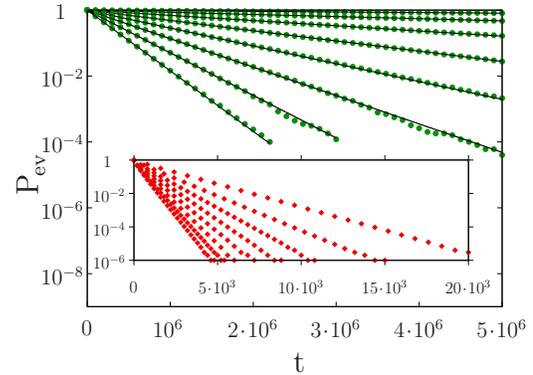}}
\caption{Survival probability of an evasive prey on a square lattice with $V = 10^4$ sites. The curves from top to bottom present the behavior of $P_{\rm ev}(t)$ when the density of the predators is $\rho = 0.002, 0.003, \ldots 0.009$. Symbols denote results of
numerical simulations and solid lines represent
an exponential fit $\exp( - b(\rho) t)$.
Inset shows the corresponding survival probability of an immobile prey.}
  \label{cartoon3}
\end{figure}

\begin{figure}[ht]
  \centerline{\includegraphics*[width=0.4\textwidth]{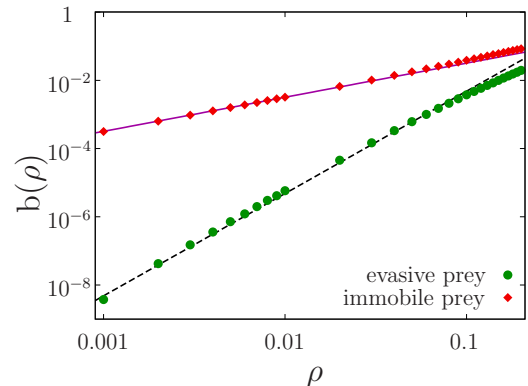}}
\caption{Characteristic decay rates $b(\rho)$ as a function of $\rho$. Diamonds represent numerical data for an immobile prey, solid line is an analytic result given by \eqref{survival2} with $\alpha =1/G$, where $G$ is given by \eqref{g}. Circles represent numerical data for an evasive prey, whereas the dashed line is a fit $b(\rho) = 4.82 \, \rho^3$.}
  \label{cartoon7}
\end{figure}

We performed Monte Carlo simulations on a $100 \times 100$ square
lattice with different numbers of predators.
In Fig.~\ref{cartoon3}, we plot the
survival probability $P_{\rm ev}(t)$ of an evasive prey
and compare it with the survival probability $P_{\rm imm}(t)$
of an \textit{immobile} prey (inset). Both display an exponential decay with time.
The comparison reveals a pronounced
difference (several orders of magnitude for sufficiently low density of predators)
in the values of respective survival probabilities.
In Fig.~\ref{cartoon7}, we plot the numerical data for the characteristic decay rates $b(\rho)$ as a function of $\rho$. This figure shows that for an immobile prey $b(\rho)$ is a linear function of $\rho$, $b(\rho) \sim \rho$, as it should be, whereas for the lazy evasive prey $b(\rho) \sim \rho^3$. This very different scaling explains why the survival probability of evasive prey is so much higher when $\rho \ll 1$.

In the following sections, we explain these findings. First, we consider
an immobile prey and evaluate an exact expression for $P_{\rm imm}(t)$ for finite lattices and in the thermodynamic limit. Then, we look on this problem from a different point of view and
obtain $P_{\rm imm}(t)$ by using a heuristic approach
 that reproduces correctly the temporal behavior of $P_{\rm imm}(t)$ and gives a correct dependence of the characteristic decay times on $V$ and other pertinent parameters, but predicts slightly different values for numerical factors.
Next, we show that in order for the evasive prey to be captured, the predators must first appear in a special configuration in which the prey will be under a direct attack but cannot evade. We specify such ``stalemate'' configurations and extend our heuristic approach to estimate $P_{\rm ev}(t)$.

\section{Survival of an immobile prey}

To set up the scene, we start with a reminder on the
so-called ``target'' problem \cite{target}, which describes the
survival of an immobile prey in presence of predators performing
independent RWs. This is an exactly solvable problem, but
existing theoretical descriptions are focused on the infinite systems.
We determine $P_{\rm imm}(t)$ for finite $V$ by
taking advantage of some already known results.

Let $\{X_n(t)\}$ denote a given realization of the trajectory of the  $n^{\rm th}$ predator
within the time interval $[0,t]$ and $X_n(\tau)$ is its position at time $\tau \in [0,t]$.
We suppose here that $\tau$ and $t$ are integers.

An indicator function ${\cal P}(t)$ of the event - that within
the time interval $[0,t]$ neither of the $N$ predators has visited the location of the prey - can be written down as
\begin{equation}
\label{indic}
{\cal P}(t) = \prod_{n = 1}^N {\cal P}\left(\{X_n(t)\}\right),
\end{equation}
where
${\cal P}\left(\{X_n(t)\}\right) = 1$ if $X_n(\tau) \neq 0$ for any $\tau \in [0,t]$ and $0$ otherwise.
Averaging \eqref{indic} we obtain\footnote{Actually
the factor before the sum is equal to $1/(V - 1)$; since we are interested in large lattices,  $V \gg 1$, we always write $V$ instead of $V-1$.}
\begin{equation}
 P_{\rm imm}(t) = \left(\frac{1}{V} \sum_{X, X \neq 0} L\left(X,t\right) \right)^N,
\end{equation}
where $L\left(X,t\right)$ is the probability that a RW, starting at site $X$ at time $0$, has not yet visited the origin at time $t$.
Explicitly, $L\left(X,t\right)$ is defined (see, e.g., \cite{hil}) as
\begin{equation}
\label{L}
 L\left(X,t\right) = \frac{1}{2 \pi i} \oint \frac{d z}{z^{t + 1}} \frac{1}{1 - z} \left[1 - \frac{G(X,z)}{G(0,z)}\right],
\end{equation}
where the integral is around the origin of the z plane and $G(X,z)$  is the lattice Green (generating) function
of the probability $G(X,t)$ to find a predator at site $X$ at time $t$. From \eqref{L} we infer (see \cite{hil,structure}) that
\begin{equation}
 \sum_{X, X \neq 0}  L\left(X, t\right) = V  - S(t),
\end{equation}
and consequently,
\begin{equation}
\label{hh}
 P_{\rm imm}(t) = \left(1 - \frac{1}{V} S(t)\right)^N.
\end{equation}

In the thermodynamic limit, $N\to\infty$ and $V\to\infty$ with $\rho=N/V$ kept constant,
equation \eqref{hh}  reduces to \eqref{survival}.
For finite lattices, the large time behavior of $S(t)$ is well-known (see e.g., refs.\cite{hil,weiss}). One finds that at sufficiently large times
\begin{equation}
\label{g}
S(t) \sim V \left[1 - e^{- t/V G}\right], \,
G =
\displaystyle \left\{\begin{array}{ll}
\displaystyle \frac{2 V}{\pi^2}, & d = 1, \\
\displaystyle \frac{\ln(c V)}{\pi}, & d = 2,
\end{array}
\right.
\end{equation}
where $c$ is a constant. For a square lattice,  $c \approx 1.8456$. For $d > 2$, $G$ is a constant as mentioned in the introduction. Overall, we recover \eqref{survival2} with $\alpha = 1/G$.

We close this section with a remark
that the target problem is closely related
to the so-called narrow escape problem,
 which arises in a number of biological processes such as
biochemical reactions in cellular microdomains (see, e.g. \cite{1,3,2}).
Here, particles (ions, molecules, proteins, etc.) move randomly in a
bounded domain (cell, compartment) enclosed by a boundary that is
perfectly reflecting everywhere,
except for a small window through which particles can escape.
Within this context, one is interested to calculate
the first passage time density, which is defined
as the time derivative of the survival probability of a target placed at  a position
of the escape window in the presence of a particle diffusing in a bounded domain.

In a $2d$ circular domain of \textit{area} $V$ with reflecting walls
containing a small escape window of size $a$ (size of the target),
the survival probability of such an immobile target in the presence of $N$
particles diffusing with
diffusion coefficient $D$ obeys (see \cite{3}), for $V \gg a^2$,
\begin{equation}
\label{n}
 P_{\rm imm}(t) \sim \exp\left(- \rho \frac{\pi D t}{\ln(V/a^2)} \right).
\end{equation}
To match $P_{\rm imm}(t)$ given by \eqref{n} and our discrete-space calculations,
we set $D= 1/4$ and notice that the characteristic time of
the decay in \eqref{n} is four times larger that the one defined by \eqref{survival2}
with $\alpha = 1/G$ determined by \eqref{g}, meaning that
an immobile prey located at the origin of a periodic
lattice will typically be caught sooner than an immobile prey hiding at a reflective boundary.
The corresponding characteristic time may be even larger if the boundary
is not smooth and the prey is located in the corner or near a cusp (see \cite{3}).
Note, as well, that the target size $a$ in \eqref{n} may be interpreted here as the sighting range of the predators; hence, \eqref{n} signifies that in 2d, dependence of the characteristic decay time on the predators' sighting range is logarithmically weak.

\section{Survival of an immobile prey revisited}
\label{revision}

We suppose from now on that $t$ is a continuous variable, and predators perform
continuous-time RWs
with a pausing-time density $\Psi(t) = \exp( - t) $.
Such a choice is for convenience only and would not cause any difference in the large-$t$ behavior
compared to
discrete time RWs.

Let $T_1$ be a random variable defining the time within the interval $[0,t]$ when the origin has been
occupied by \textit{one or more predators} simultaneously.
Then,
 ${\cal P}(t)$ in \eqref{indic} reads
\begin{equation}
{\cal P}(t) =
\displaystyle \left\{\begin{array}{ll}
\displaystyle 1, & T_1 \equiv 0, \\
\displaystyle 0, & T_1 >  0.
\end{array}
\right.
\end{equation}
Hence, once we know the probability density  $P(T_1)$,
we may find the survival probability $P_{\rm imm}(t)$ of an immobile target from
$P_{\rm imm}(t) = P(T_1 = 0)$.

Occupancy time $T_1$, i.e., a time spent on a given lattice site by one (or simultaneously $k$) random walkers,
have been studied in a number of works (see, e.g. refs.\cite{katja,beresh,N}).
An important for us result has been conjectured and
verified numerically in ref.\cite{beresh}; it states  that  $P(T_1)$ is a
Gaussian
\begin{equation}
P(T_1) \sim \exp\left(- \frac{(T_1 - \overline{T_1})^2}{2 \sigma_1^2}\right)
\end{equation}
and thus is entirely defined by the first two moments of $T_1$: mean value $\overline{T_1}$ and the variance $\sigma_1^2 = \overline{T_1^2} - \overline{T_1}^2$. Evidently,
$\overline{T_1}$ grows linearly with $t$, and hence $T_1 = 0$ is a
large deviation from the most probable value, such that, in principle,
an assumption that $P(T_1)$ is Gaussian may not be valid in this domain.
We proceed to show, however, that an estimate
\begin{equation}
\label{estimate}
 P_{\rm imm}(t) \sim P(T_1 = 0) \sim \exp\left(- \frac{\overline{T_1}^2}{2 \sigma_1^2}\right),
\end{equation}
reproduces correctly the dependence of $P_{\rm imm}(t)$ on time and other parameters;
only numerical factors in the characteristic decay times are incorrect.

Let $\Psi_0(\tau)$ be the indicator function of the event that at time moment $\tau$ none of $N$ predators is at the origin:
\begin{equation}
\label{ind}
\Psi_0(\tau) = \prod_{n = 1}^{N} \left[ 1 - I\left(X_n(\tau)\right)\right], \, I(X) =
\displaystyle \left\{\begin{array}{ll}
\displaystyle 1, & X = 0 \\
\displaystyle 0, & X \neq  0.
\end{array}
\right.
\end{equation}
Therefore, random variable $T_1$ is explicitly defined by
\begin{equation}
 T_1 = \int^t_0 d\tau \, \left(1 - \Psi_0(\tau)\right).
\end{equation}
The first moment of $T_1$ is now calculated to give
\begin{equation}
 \label{lu}
 \overline{T_1} =  \int^t_0 d\tau \, \left[1 - \left(1 - \frac{z_{\tau}}{V} \right)^N\right], \, z_{\tau}  = 1 - G_{\tau},
\end{equation}
where $G_{\tau} = G_{\tau}(0)$ is the probability that a RW,
commencing at the origin, is at the origin at
time moment $\tau$.

For the variance of $T_1$ we find
\begin{eqnarray}
\label{qq}
\sigma^2_1 = \int^t_0 d\tau_1 \int^t_{0} \, d\tau_2 \, \, \phi_{0,0}(\tau_1,\tau_2),
\end{eqnarray}
where $\phi_{0,0}(\tau_1,\tau_2)$ is the two-time correlation function:
\begin{eqnarray}
\phi_{0,0}(\tau_1,\tau_2)  &=& \overline{\Psi_0(\tau_1) \, \Psi_0(\tau_2)} - \overline{\Psi_0(\tau_1)} \cdot \overline{\Psi_0(\tau_2)} \nonumber\\
&=& \left(1 - \frac{z_{\tau_1} + z_{\tau_2} - G_{|\tau_1 - \tau_2|} z_{\tau_2}}{V} \right)^N \nonumber\\
&-& \left(1 - \frac{z_{\tau_2}}{V}\right)^N \left(1 - \frac{z_{\tau_1}}{V}\right)^N.
\end{eqnarray}

Consider first the behavior predicted by \eqref{estimate} in
the thermodynamic limit. We find that
\begin{equation}
\label{qqqq}
 \phi_{0,0}(\tau_1,\tau_2) \sim \rho \, G_{|\tau_1 - \tau_2|},
\end{equation}
implying that the occupation
of the origin (or any other lattice site) of an infinite lattice
is a stationary process with long-range \textit{algebraic} correlations.
Curiously enough, this behavior resembles very much the behavior observed experimentally for ``blinking" of nanoscale light emitters \cite{eli}.

Integrating \eqref{qqqq} and noticing that $\overline{T_1} \sim \rho \, t$, we obtain
\begin{equation}
\frac{\overline{T_1}^2}{2 \sigma_1^2} \sim \rho
 \left\{\begin{array}{lll}
\displaystyle f_1 \, (8 t/\pi)^{1/2}, & d = 1, \\
\displaystyle f \, \pi t/\ln(t), & d =  2,\\
\displaystyle f \, t/G', & d >  2,
\end{array}
\right.
\end{equation}
where $f_1 = 3 \pi/32$, $f = 1/4$ and $G' = \int^{\infty}_0 d\tau \, G_{\tau}$. Note that $G'$ has exactly the same meaning
as $G$ in \eqref{g}, namely it equals the mean number of visits to the origin
of an infinite lattice by a continuous-time random walk within an infinite time.

For finite but large $V$, from equations \eqref{lu} and \eqref{qqqq} we find that in the leading in $1/V$ order
\begin{equation}
\label{qu}
\overline{T_1} \sim  \rho \, t, \,\,\, \sigma_1^2 \sim 2 \rho \int^t_0 d\tau_1 \int^{\tau_1}_0 d\tau_2 \, \left(G_{\tau_1 - \tau_2} - \frac{1}{V}\right),
\end{equation}
which yields
\begin{equation}
\frac{\overline{T_1}^2}{2 \sigma_1^2} \sim \frac{\rho t}{4}
 \left\{\begin{array}{lll}
\displaystyle \pi^2/2 V, & d = 1, \\
\displaystyle \pi/\ln(V), & d =  2,\\
\displaystyle 1/G', & d >  2.
\end{array}
\right.
\end{equation}
Hence, both
in the thermodynamic limit and for finite $V$
our approximate approach predicts
correct (non trivial in low dimensions)
temporal evolution of $P_{\rm imm}(t)$, correct (non-trivial) dependence on $V$, but overestimates numerical factors in the characteristic decay times.

\section{Survival of a lazy evasive prey}

\begin{figure}[ht]
  \centerline{\includegraphics*[width=0.3\textwidth]{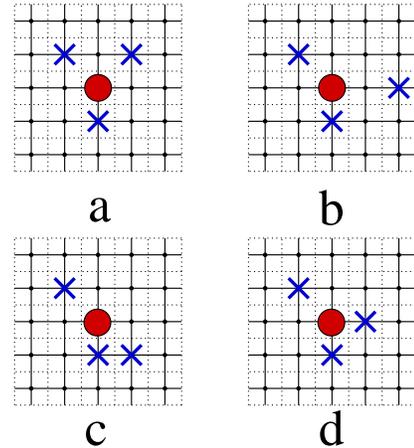}}
\caption{Four basic stalemate configurations from which the lazy prey cannot escape; the prey will be captured on the next step.}
  \label{cartoon2}
\end{figure}

We now can explain why the behavior of the survival probability
becomes so markedly different when the prey is evasive. First, we note the obvious
fact that the evasive prey cannot be captured by a single predator.
On a square lattice, the evasive prey cannot be captured by
two predators, i.e., a more collective effort is required.
In order  to catch the prey, the predators have to create
a stalemate type situation in which the prey can be
attacked but cannot evade. On the square lattice, three predators can do this job.
Several such configurations
created by three predators are depicted in Fig.~\ref{cartoon2}.

When $\rho \ll 1$,  we may consider only
lowest order ``reaction events'' which involve just
three predators, as depicted in Fig.~\ref{cartoon2}, and disregard
higher order configurations, involving four and more predators.

\begin{figure}[ht]
  \centerline{\includegraphics*[width=0.4\textwidth]{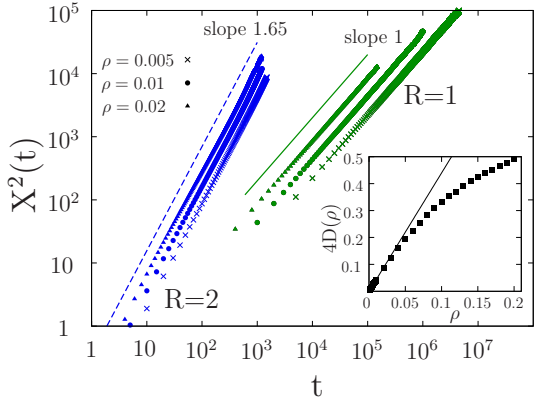}}
\caption{Mean-squared displacement of an evasive prey till the moment of its capture. Shown are simulation results for the predator sighting ranges $R = 1$ and $R = 2$. (Inset) The effective diffusion coefficient (determined via relation $4 D(\rho) = \overline{X^2(t)}/t$) of the prey for $R = 1$. The solid line in the inset represents $D(\rho) \sim \rho$.}
  \label{cartoon5}
\end{figure}

Due to the lazy evasion, the prey effectively performs  a
diffusive-type motion. A rigorous proof of this statement could be extremely challenging, and even quantifying the motion is not trivial, as the prey is eventually caught. If we limit ourselves to the realizations when the lazy prey has not been caught up to time $t$ and consider the mean-squared displacement of the prey, we find that it grows linearly with time (see Fig.~\ref{cartoon5}). Its diffusion coefficient $D \sim \rho$ (inset in Fig.~\ref{cartoon5}) and is small compared with the diffusion coefficient of the predators (which is $D_p = 1/4$, according to our definition of the motion of the predators). There is also another indirect argument to ignore the motion of the prey.
In low-dimensional systems, the leading asymptotical behavior
of the survival probability of the diffusive prey in presence of diffusive predators is independent of the diffusion coefficient of the prey \cite{bray,we} as well as of the reaction probability \cite{structure}.
Hence we neglect diffusive-type motion of the prey and suppose that it is
immobile.

Thus, the most significant effect of the lazy evasion tactics is the change of the effective reaction order --- the prey can be captured only when any three predators appear
in a stalemate configuration. Summing up these arguments,
we estimate $P_{\rm ev}(t)$ of an evasive prey as
\begin{equation}
\label{AC}
\ln P_{\rm ev}(t) \sim A \, C \, \ln P_{3}(t).
\end{equation}
Here $C = 28$ is the total number of stalemate configurations, see Fig.~\ref{cartoon2}, and $P_{3}(t)$ is the survival probability of an \textit{immobile} prey which is captured only by three (or more) predators; that is,
the prey can coexist with one or two predators and it is caught when
three predators appear simultaneously at the site it occupies. The numerical factor $A$ in \eqref{AC} is unknown, it emerges since we approximately consider any stalemate configuration involving three predators as a simultaneous encounter of predators at the same site. (One can also envision that stalemate configurations have different weights.) We will extract the value of $A$ from the fit to numerical data.

Following our heuristic interpretation of the survival probability of an immobile prey,
we estimate $P_3(t)$ as
\begin{equation}
P_3(t) \sim \exp\left(- \frac{\overline{T_3}^2}{2 \sigma^2_3}\right),
\end{equation}
where now $T_3$ is a
random variable equal to the fraction of time within the interval $[0,t]$ when the origin has been occupied by at least three predators \textit{simultaneously}.

We define next the indicator functions:
\begin{eqnarray}
\Psi_1(\tau) &=& \sum_{l = 1}^{N} I\left({X}_l(\tau)\right) \prod_{n = 1, n \neq l}^{N} \left[ 1 - I\left(X_n(\tau)\right)\right],
\end{eqnarray}
and
\begin{eqnarray}
\Psi_2(\tau) &=& \sum_{l = 1}^N \sum_{p = 1, p \neq l}^{N} I\left({X}_l(\tau)\right) \, I\left({X}_p(\tau)\right) \times \nonumber\\ &\times& \prod_{n = 1, n \neq l, p}^{N} \left[ 1 - I\left(X_n(\tau)\right)\right],
\end{eqnarray}
of the events that
at time moment $\tau$ there is just one or just two predators at the origin, respectively.
Then, $T_3$ is given by
\begin{equation}
  T_{3} =  \int^t_0 d\tau \, \left[1 - \Psi_0(\tau) - \Psi_1(\tau) - \Psi_2(\tau)\right],
\end{equation}
and its first two moments obey:
\begin{eqnarray}
 \overline{T_3} &=& \int^t_0 d\tau \, \Big(1 - \left(1 - \frac{z_{\tau}}{V}\right)^N - \frac{N z_{\tau}}{V} \left(1 - \frac{z_{\tau}}{V}\right)^{N - 1} - \nonumber\\
&-& \frac{1}{2} \frac{N (N - 1) z_{\tau}^2}{V^2} \left(1 - \frac{z_{\tau}}{V}\right)^{N - 2}\Big),
\end{eqnarray}
and
\begin{equation}
\sigma_3^2 = \overline{T_3^2} - \overline{T_3}^2 =  \int^t_0 d\tau_1 \int^t_0 d \tau_2 \, \sum_{i,j = 0}^2 \, \phi_{i,j}(\tau_1,\tau_2),
\end{equation}
where the two-time correlation functions are defined by
\begin{equation}
\phi_{i,j}(\tau_1,\tau_2) = \overline{\Psi_{i}(\tau_1) \Psi_{j}(\tau_2)}  - \overline{\Psi_{i}(\tau_1)} \cdot \overline{\Psi_{j}(\tau_2)}.
\end{equation}
In the leading in $N/V$ order, $\overline{T_3}$ is simply
\begin{equation}
\overline{T_3} \sim \rho^3 \, t.
\end{equation}
For finite lattices,
the two-time correlation functions follow
\begin{equation}
 \phi_{i,j}(\tau_1,\tau_2) \sim a_{i,j}(\rho) \left(G_{|\tau_1 - \tau_2|} - \frac{1}{V}\right),
\end{equation}
where $a_{i,j}(\rho)$ are polynomials in $\rho$. After straightforward, but tedious calculations, we find that in the leading in $N/V$ order $\sum_{i,j} a_{i,j}(\rho) \sim \rho^3$ and hence, $\overline{T_3} \sim \rho^2 \overline{T_1}$ and $\sigma_3^2 \sim \rho^2 \sigma_1^2$. This implies that for sufficiently large times, the survival probability $P_3(t)$ of an immobile prey which can be captured when any three predators appear simultaneously on the site it occupies, and the survival probability $P_{\rm imm}(t)$ of the target problem are related to each other through
\begin{equation}
\ln P_3(t) \sim  \rho^2 \ln P_{\rm imm}(t).
\end{equation}
Consequently, for finite square lattices $P_{\rm ev}(t)$ of an evasive prey obeys
\begin{equation}
\label{ev}
 \ln P_{\rm ev}(t) \sim A \, C \rho^2 \ln P_{\rm imm}(t) \sim - A \, C \, \rho^3 \frac{\pi t}{\ln(c V)}.
\end{equation}
From the fit to numerical data,
we get $A \approx 0.55$.

In a similar fashion, we find that for an infinitely large square lattice
\begin{equation}
\label{ev2}
 \ln P_{\rm ev}(t)  \sim - A \, C \, \rho^3 \frac{\pi t}{\ln(t)}.
\end{equation}
Equations \eqref{ev}--\eqref{ev2} lead to the announced result \eqref{survival3}.

\section{Discussion} We studied the survival of a
prey in the presence of predators performing RWs on sites of a
square lattice. We analyzed a minimal-effort evasion tactics
in which the prey tries to avoid predators by
stepping away whenever a predator
appears on one of the neighboring sites; otherwise the prey stays still. We showed that this strategy leads to the great enhancement of the survival probability in comparison with the stay-still strategy \cite{pascal2,pascal1}. More precisely, when the density $\rho$ of the predators is small, the life expectancy of the immobile prey scales as $\rho^{-1}$ whereas the minimal-evasion strategy results in the life expectancy of the order of $\rho^{-3}$.

Several interesting additional conclusions are as follows:

a) We assumed that the predators perform conventional RWs until the prey appears within their sighting ranges. Within this picture, the characteristic relaxation time of the survival probability of both an evasive and an immobile prey
appears to be proportional to the factor $G$, which defines the mean number of returns to the origin within an infinite time during which a predator commences its random motion at the origin.
In the case of a search for the prey by conventional RWs, in two dimensions $G$ is large when $V$ is large, because $G \sim \ln(V)$ but attains a finite value (as $V \to \infty$) for $d > 2$. Such a behavior is associated with the fact that the spatial dimension $d=2$ coincides with the fractal dimension of RWs.
Playing on the side of the predators, it becomes clear that more efficient search for the prey in two dimensions
will be realized when the predators' trajectories have
a fractal dimension $< 2$ (i.e. the best option would be to perform
ballistic motion).

b) We focused on a particular situation (the minimal model)
in which the sighting range of both the prey ($r$)
and of the predators ($R$)
were equal to one lattice spacing.
When $R > r$, e.g. $R = 2$ and $r=1$, we observed a different behavior: After the prey notices the predator, it can never escape from the sighting range of this predator.
In this situation, as time increases, more and more predators turn
from a random-search phase
to the phase of the direct chase, so that the prey is accompanied
by a tail of chasing predators. In our model, neither of the species
is superior to the other with respect to speed and hence
the predators who are directly chasing the prey are harmless - they just follow the prey but can never catch it, as long as the prey is not caged. We observed that in some realizations of the prey-predator contests on \textit{sufficiently
small} lattices, all $N$ predators were directly chasing the prey which, however, survives to eternity.
On the other hand, the predators who perform a direct chase: (i) block some possible directions for escape, such that it suffices now to meet just one randomly searching predator in order to create a stalemate configuration and,
(ii) exert some pressure on the prey prompting it to move almost ballistically.
Encounters with the predators who have not yet started a direct chase
suppresses pure ballistic motion, but still the prey performs effectively a \textit{superdiffusive} motion, in contrast to the case when $R = r$. Here, the mean-square displacement $\overline{X^2(t)}$ of the prey,
until the moment it is caught, obeys $\overline{X^2(t)} \sim t^z$ with $z \approx 1.65$ (see Fig.\ref{cartoon5}, $R = 2$ case) meaning that contrary to predators, whose random motion is recurrent in $2d$ and whose $S(t)$ grows
sublinearly with time (due to a logarithmic correction), the
fractal dimension of the prey's trajectories $d_w \approx 1.2 < 2$, and the mean number of
distinct sites it visits grows linearly with $t$. Thus, paradoxically, although the predators who are in a direct chase phase are harmless to the prey, they force it to explore more new sites and, consequently, to die sooner.
This interesting case bears further investigation.

c) There should be an optimal value of the prey sighting range $r$.
Clearly, a sighting range which is too small would not
result in an effective evasion. On the other hand, having a too-large $r$ is not good either.
If $r$ exceeds the mean distance between the predators, the prey will always see
predators attempting a move
and thus
would most likely stay still.

d) Finally, we remark that our results shed some light on an interesting problem of first passage times $\tilde{t}$
to rare density fluctuations in diffusive systems. Within this context, one is interested to calculate the first passage time density $Q^{(k)}(\tilde{t})$
of the event when $k$ out of $N$ particles performing RWs on a lattice
appear for the first time $\tilde{t}$ simultaneously
on a specific lattice site \cite{sanders}. Our analysis suggests
that for sufficiently large $\tilde{t}$ the first passage time density obeys $\ln Q^{(k)}(\tilde{t}) \sim - \rho^k \tilde{t}/G$, with $G$ defined by \eqref{g}.

\begin{acknowledgments}
We wish to thank Ted Cox for inspiring our interest in this subject. We also acknowledge helpful discussions with Olivier
B\'enichou, Raphael Voituriez and Alexander Berezhkovskii on
many aspects of sojourn and
first passage times distributions in confined systems.  G.O. is partially supported by Agence Nationale de la Recherche
(ANR) under grant ``DYOPTRI - Dynamique et Optimisation des Processus de
Transport Intermittents''. P.L.K. is grateful for financial support
from NSF grants CHE-0532969 and CCF-0829541.
P.L.K. and J.K. acknowledge hospitality of Laboratoire de Physique Th\'eorique
 et Mod\`eles Statistiques, Universit\'e Paris 11, and of Laboratoire de Physique de la Mati\`ere Condens\'ee, Ecole Polytechnique.
\end{acknowledgments}



\end{article}

\begin{thebibliography}{99}


\bibitem{nahin}
Nahin P (2007)
Chases and escapes: The mathematics of pursuit and evasion.
{\it Princeton University Press, Princeton NJ}.
%
\bibitem{webb}
Weihs D, Webb P W (1984) Optimal avoidance and evasion tactics in
predator-prey interactions. {\it J.  Theor.  Biol.} {\bf 106}: 189--206.
%
\bibitem{miller}
Miller G F, Cliff D (1994)
Co-evolution of pursuit and evasion: Biological and Game-theoretical foundations.
{\it Technical Report, School of Cognitive and Computing Sciences, University of Sussex}.
%
\bibitem{pascal0}
Krapivsky P L, Redner S (1996)
Kinetics of a diffusive capture process: lamb besieged by a pride of lions.
{\it J. Phys. A} {\bf 29}: 5347--5357;
Redner S, Krapivsky P L (1999)
Capture of the lamb: Diffusing predators seeking a diffusing prey.
{\it Amer. J. Phys.} {\bf 67}: 1277--1283.
%
\bibitem{isaacs} Isaacs R (1965) Differential Games: A Mathematical Theory with Applications to Warfare and Pursuit, Control and Optimization. {\it John Wiley \and Sons, New York}.
%
\bibitem{pascal2}
Bray A J, Majumdar S N,  Blythe R A (2003) Formal solution of a class of reaction-diffusion models: Reduction to a single-particle problem.
{\it Phys  Rev  E} {\bf 67}: 060102.
%
\bibitem{pascal1}
Moreau M, Oshanin G, B\'enichou O, Coppey M (2003) Pascal principle for diffusion-controlled trapping reactions. {\it Phys.  Rev.  E} 67: 045104;
(2004) Lattice theory of trapping reactions with mobile species.
{\it Phys.  Rev.  E} 69: 046101.
%
\bibitem{target}
Tachiya M (1983) Theory of diffusion-controlled reactions: Formulation of the bulk reaction rate in terms of the pair probability.
{\it Radiat.  Phys.  Chem.} {\bf 21}: 167--175;
Blumen A, Zumofen G, Klafter J (1984) Target annihilation by random walkers.
{\it Phys.  Rev.  B} {\bf 30}: 5379--5382;
Redner S, Kang K (1984) Kinetics of the 'scavenger' reaction.
{\it J.  Phys.  A} {\bf 17}: L451--L455.
%
\bibitem{dvor}
Dvoretzky A, Erd\"os P (1951) Some problems on random walks in space.
In: Proceedings of the Second Berkeley Symposium on
Mathematical Statistics and Probability, Edited by J. Neyman,
{\it Berkeley, University of California Press} 353--367.
%
\bibitem{hil}
Brummelhuis M J A M, Hilhorst H J (1991) Covering of a finite lattice by a random walk.
{\it  Physica  A} {\bf 176}: 387--408.
%
\bibitem{structure}
B\'enichou O, Moreau M, Oshanin G (2000) Kinetics of stochastically gated diffusion-limited reactions and geometry of random walk trajectories.
{\it Phys.  Rev.  E} {\bf 61}: 3388--3406.
%
\bibitem{weiss}
Weiss G H, Havlin S, Bunde A (1985) On the survival probability of a random walk in a finite lattice with a single trap.
{\it J.  Stat.  Phys.} {\bf 40}: 1572--9613.
%
\bibitem{1}
Grigoriev I V, Makhnovskii Y A, Bereshkovskii A M, Zitserman V Y (2002) 	 
Kinetics of escape through a small hole.
{\it J. Chem. Phys.} 116: 9574--9577.
%
\bibitem{3}
Schuss Z, Singer A, Holcman D (2007) The narrow escape problem for diffusion in cellular microdomains.
{\it Proc. Natl. Acad. Sci. USA} {\bf 104}: 16098--16103.
%
\bibitem{2}
B\'enichou O, Voituriez R (2008) Narrow-escape time problem: Time needed for a particle to exit a confining domain through a small window
{\it Phys.  Rev.  Lett.} {\bf 100}: 168105.
%
\bibitem{katja}
Weiss G H, Shuler K E, Lindenberg K (1983) Order statistics for first passage times in diffusion processes.
{\it J. Stat. Phys.} {\bf 31}: 255--278.
%
\bibitem{beresh}
Bogun\`a M, Berezhkovskii A M,  Weiss G H (2000) Occupancy of a single site by many random walkers.
{\it Phys. Rev. E} 62: 3250--3256.
%
\bibitem{N}
B\'enichou O, Coppey M, Klafter J, Moreau M, Oshanin G (2003)  	On the joint residence time of N independent two-dimensional Brownian motions.
{\it J.  Phys.  A} 36: 7225--7231.
%
\bibitem{eli}
Stefani F D, Hoogenboom J P, Barkai E (2009) Beyond quantum jumps: Blinking nanoscale light emitters. {\it Physics Today} February issue: 34--39.
%
\bibitem{bray}
Bray A J, Blythe R A (2002) 	
Exact Asymptotics for One-Dimensional Diffusion with Mobile Traps.
{\it Phys.  Rev.  Lett.} {\bf 89}: 150601;
Blythe R A,  Bray A J (2003) Survival probability of a diffusing particle in the presence of Poisson-distributed mobile traps.
{\it Phys.  Rev.  E} {\bf 67}: 041101.
%
\bibitem{we}
Oshanin G, B\'enichou O, Coppey M, Moreau M (2002) Trapping reactions with randomly moving traps: Exact asymptotic results for compact exploration.
{\it Phys.  Rev.  E} {\bf 66}: 060101;
Yuste S B, Oshanin G, Lindenberg K, B\'enichou O, Klafter J (2008) Survival probability of a particle in a sea of mobile traps: A tale of tails.
{\it Phys. Rev. E} {\bf 78}: 021105.
%
\bibitem{sanders}
Sanders D P, Larralde H (2008) 	How rare are diffusive rare events?
{\it Europhys. Lett.} {\bf 82}: 40005--40011.
%
\end{thebibliography}
\end{document}